\documentclass[letter,aps,showpacs,superscriptaddress,nofootinbib,tightenlines,twocolumn,prl]{revtex4}

\usepackage{amsfonts}
\usepackage{multirow}
\usepackage{mathrsfs}
\usepackage{graphicx}
\usepackage{amsmath}
\usepackage{amssymb}
\usepackage{bm}
\usepackage{bbm}
\usepackage{color}
\usepackage{slashed}
\usepackage{diagbox}
\usepackage{ulem}
\def\OMIT#1{}

\newcommand{\nn}{\nonumber}

\newcommand{\beq}{\begin{equation}}
\newcommand{\eeq}{\end{equation}}
\newcommand{\bqa}{\begin{eqnarray}}
\newcommand{\eqa}{\end{eqnarray}}

\begin{document}
%\preprint{}
%%%%%%%%%%%%%%%%%%%%%%%%%%%%%%%%%%%%%%%%%%%%%%%%%%%%%%%%%%%%%%%%%%%%%%%%%%%%%%
\title{\mbox{}\\[10pt]
Mixed electroweak-QCD corrections to $e^+e^-\to HZ$ at Higgs factory}

%%%%%%%%%%%%%%%%%%%%%%%%%%%%%%%%%%%%%%%%%%%%%%%%%%%%%%%%%%%%%%%%%%%%%%%%%%%%%%

\author{Qing-Feng Sun~\footnote{qfsun@mail.ustc.edu.cn}}
\affiliation{Department of Modern Physics, University of Science and Technology of China, Hefei, Anhui 230026,
China\vspace{0.2cm}}
\affiliation{Institute of High Energy Physics and Theoretical Physics Center for
 Science Facilities, Chinese Academy of
 Sciences, Beijing 100049, China\vspace{0.2cm}}

\author{Feng Feng~\footnote{F.Feng@outlook.com}}
\affiliation{China University of Mining and Technology, Beijing 100083, China\vspace{0.2cm}}
\affiliation{Institute of High Energy Physics and Theoretical Physics Center for
Science Facilities, Chinese Academy of
Sciences, Beijing 100049, China\vspace{0.2cm}}

\author{Yu Jia~\footnote{jiay@ihep.ac.cn}}
\affiliation{Institute of High Energy Physics and Theoretical Physics Center for
Science Facilities, Chinese Academy of
Sciences, Beijing 100049, China\vspace{0.2cm}}
\affiliation{School of Physics, University of Chinese Academy of Sciences, Beijing 100049,
China\vspace{0.2cm}}
\affiliation{Center
for High Energy Physics, Peking University, Beijing 100871,
China\vspace{0.2cm}}

\author{Wen-Long Sang~\footnote{wlsang@ihep.ac.cn}}
 \affiliation{School of Physical Science and Technology, Southwest University, Chongqing 400700, China\vspace{0.2cm}}

\date{\today}

%%%%%%%%%%%%%%%%%%%%%%%%%%%%%%%%%%%%%%%%%%%%%%%%%%%%%%%%%%%%%%%%%%%%%%%%%%%%%%
\begin{abstract}
The prospective Higgs factories, exemplified by ILC, FCC-ee and CEPC,
plan to conduct the precision Higgs measurements at the $e^+e^-$ center-of-mass energy around 250 GeV.
The cross sections for the dominant Higgs production channel, the Higgsstrahlung process, can be measured to a
(sub-) percent accuracy. Merely incorporating the well-known next-to-leading order (NLO) electroweak corrections appears
far from sufficient to match the unprecedented experimental precision.
In this work, we make an important advancement toward this direction by investigating the mixed electroweak-QCD corrections to
$e^+e^-\to HZ$ at next-to-next-to-leading order (NNLO) for both unpolarized and polarized $Z$ boson.
The corrections turn out to reach one percent level of the Born-order results, thereby must be incorporated in the future confrontation with the data.
\end{abstract}
%%%%%%%%%%%%%%%%%%%%%%%%%%%%%%%%%%%%%%%%%%%%%%%%%%%%%%%%%%%%%%%%%%%%%%%%%%%%%%
\pacs{\it 12.15.Lk, 12.38.-t, 13.66.Fg, 14.80.Bn}

%%%%%%%%%%%%%%%%%%%%%%%%%%%%%%%%%%%%%%%%%%%%%%%%%%%%%%%%%%%%%%%%%%%%%%%%%%%%%%%%%%%%%%%%%%%%%%%%%%%%%%%%%%%%%%%%%%%%%%%%%%%%%%%%%%%%%%%%%%%%%%%%%%%%%%%%%%%
%12.15.Lk Electroweak radiative corrections
%14.80.Bn Standard-model Higgs bosons
%12.38.-t Quantum chromodynamics
%12.38.Bx Perturbative calculations
%13.66.Fg Gauge and Higgs boson production in e?e+ interactions
%12.15.Ji Applications of electroweak models to specific processes
%14.70.Hp Z bosons
%%%%%%%%%%%%%%%%%%%%%%%%%%%%%%%%%%%%%%%%%%%%%%%%%%%%%%%%%%%%%%%%%%%%%%%%%%%%%%

\maketitle

{\noindent \it \color{blue}Introduction.} The ground-breaking discovery of the 125 GeV boson at
CERN Large Hadron Collider (\textsf{LHC}) in 2012 has opened a new era in particle physics~\cite{Aad:2012tfa,Chatrchyan:2012xdj}.
It is of the highest priority to scrutinize the property of this Higgs-like boson, in order to penetrate
into the mechanism of electroweak symmetry breaking, and to seek the footprint
of new physics.
In contrast to the enormous backgrounds at \textsf{LHC}, the clean environment
renders the $e^+e^-$ collider to be a much more appealing option to
conduct precision Higgs measurements.

Recently, three next-generation $e^+e^-$ colliders have been proposed to serve as Higgs factory:
International Linear Collider (\textsf{ILC})~\cite{Baer:2013cma,Asner:2013psa}, Future Circular Collider (\textsf{FCC-ee})~\cite{Gomez-Ceballos:2013zzn},
and Circular Electron-Positron Collider (\textsf{CEPC})~\cite{CEPC-SPPCStudyGroup:2015csa,CEPC-SPPCStudyGroup:2015esa}.
All of them intend to operate at center-of-mass (CM) energy within the $240\sim 250$ GeV range,
and plan to accumulate about $10^5-10^6$ Higgs boson events. Around such energy, the Higgsstrahlung process,
$e^+e^-\to HZ$, becomes the dominant Higgs production channel, much more important than the $WW/ZZ$-fusion processes,
and the recoil mass technique can be applied to precisely measure the $HZ$ event yield
and the Higgs boson mass.
Consequently, $\sigma(e^+e^-\to HZ)$ is anticipated to be measured to an exquisite accuracy,
{\it e.g.}, 1.2\% at \textsf{ILC}, 0.5\% at \textsf{CEPC}, and 0.4\% at \textsf{FCC-ee}.
Moreover, various Higgs couplings, exemplified by $H\to gg,c\bar{c}$,
can also be precisely measured at Higgs factory, otherwise very difficult to access at \textsf{LHC}.
Furthermore, it has also been recently suggested that the $\sigma(HZ)$ could serve as a sensitive probe for various new physics scenarios~\cite{Katz:2014bha,Englert:2014uua,Craig:2014una,Huang:2015izx,Craig:2015wwr,Ge:2016zro}.

Needless to say, in order to confidently interpret the future experimental measurements,
one must develop a comprehensive knowledge on the Standard Model (SM) predictions to the Higgsstrahlung process.
The leading order (LO) prediction to this process was known long ago~\cite{Ellis:1975ap,Ioffe:1976sd,Bjorken:1977wg}.
The NLO electroweak corrections have also been
available for a while, independently addressed by three groups~\cite{Fleischer:1982af,Kniehl:1991hk,Denner:1992bc}.
For a light Higgs boson and at Higgs factory energies, the NLO weak corrections can reach a
few percent level, thereby must be incorporated in phenomenological analysis.

To match the projected sub-percent accuracy of the cross section measurements at \textsf{CEPC} and \textsf{FCC-ee},
it seems compulsory to incorporate even higher order corrections. The next most important corrections are the ${\mathcal O}(\alpha^2)$ electroweak corrections and the mixed electroweak-QCD ${\mathcal O}(\alpha\alpha_s)$ corrections.
While the former is exceedingly challenging to compute, the latter is much more tractable and may be more significant in magnitude owing to the occurrence of the QCD coupling constant. It is the very goal of this work to comprehensively investigate the ${\mathcal O}(\alpha\alpha_s)$ corrections to the Higgsstrahlung process at Higgs factory.

%-------------------------
\begin{figure}[htbp]
 	\centering
 	\includegraphics[width=0.5\textwidth]{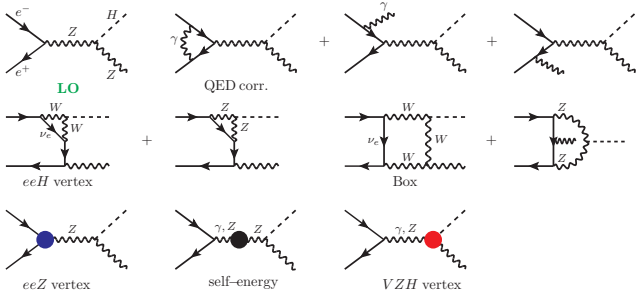}
 	\caption{LO diagram for $e^+e^-\to HZ$ and examples of QED ${\mathcal O}(\alpha)$
corrections and weak one-loop corrections, consisting of $eeH$ vertex
corrections, box diagrams, and corrections to the $eeZ$ vertex, the $\gamma/Z$
self-energy and $VZH$ vertex. The latter three types of corrections also
include ${\mathcal O}(\alpha\alpha_s)$ corrections as shown in Fig. 2.
\label{Fig:eeHZ:diagrams}}
\end{figure}
%-------------------------

%-------------------------
 \begin{figure}[htbp]
 	\centering
 	\includegraphics[width=0.5\textwidth]{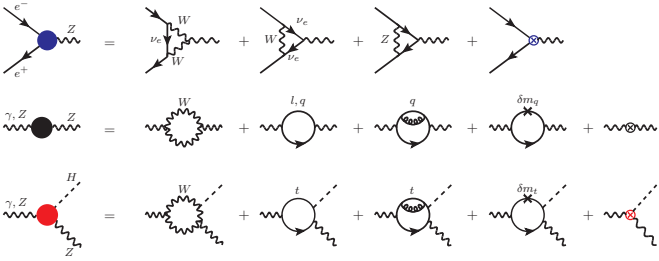}
 	\caption{Representative diagrams for the weak ${\mathcal O}(\alpha)$ and ${\mathcal O}(\alpha\alpha_s)$ corrections to the
$eeZ$ vertex, $\gamma/Z$ self-energy, and $VZH$ vertex.
 The cross represents the quark mass counterterm in QCD,
 a cap denotes the electroweak counterterm.
 \label{Fig:SE:Vertex:diagrams}}
 \end{figure}
%-------------------------

\vspace{0.3 cm}
{\noindent \it \color{blue} Leading-order results.}
By safely neglecting the electron mass owing to its exceedingly tiny Yukawa coupling,
there is only a single $s$-channel Feynman diagram for the LO Higgsstrahlung process,
as depicted in Fig.~\ref{Fig:eeHZ:diagrams}.
In the CM frame, the amplitude for $e^+(k_1,-\sigma)+e^-(k_2,\sigma)\to H(p_H)+Z(p_Z,\lambda)$ reads:
%-------------------------
\beq
%-------------------------
\label{LO:helicity:amplitude}
%-------------------------
\mathcal{M}_0^{\sigma,\lambda}=e^2 g_e^{\sigma}{M_Z \over s_W c_W}{1\over s-M_Z^2}\bar{v}(k_1)
\slashed{\varepsilon}^{*}_{\lambda} P_{\sigma}u(k_2),
%-------------------------
\eeq
%-------------------------
%$\texttt{P}_{\pm} ={1\pm\gamma^5\over 2}$
where $P_{\pm} ={1\pm\gamma^5\over 2}$ are chirality projectors, $\varepsilon_{\lambda}^\mu$
denotes the polarization vector of the $Z$ boson, with $\lambda=0 (\pm 1)$ being the longitudinal(transverse) polarization. $\sigma=\pm{1\over2 }$ represents the helicity of the incoming electron or positron (often we use the shorthand $\sigma=\pm$ for brevity).
To warrant a nonvanishing amplitude, the positron must carry the opposite helicity with respect to the electron.
We follow the conventions in~\cite{Sirlin:1980nh} to define the Weinberg angle as $c_W\equiv{M_W\over M_Z}$, and $s_W\equiv\sqrt{1-c_W^2}$. The $Zf\bar{f}$ couplings $g_f^\pm$ are defined following~\cite{Denner:1992bc}.
% couplings are given by $g_f^+=-{s_W\over c_W}Q_f$, $g_f^-= {1\over s_W c_W}(I_{f}^3 -s_W^2 Q_f)$, where $ Q_f $, $I_{f}^3 $ are the % electric charge and the third component of the weak isospin of the fermion, respectively. For electron, one substitutes $Q_e=-1$ and % $I_{e}^3 = -{1\over 2}$.

For simplicity, we will consider the unpolarized $e^+(e^-)$ beams, which is the case for
\textsf{CEPC} and \textsf{FCC-ee}.
% Upon averaging over the initial-state $e^+(e^-)$ helicities,
The LO differential cross section for polarized $Z$ then reads
%-------------------------
\begin{align}
%-------------------------
\label{LO:Differential:X:section}
& \frac{d \sigma_\lambda^{(0)}}{d{\cos\theta}}=\frac{\pi\alpha^2\beta}{16c_W^2 s_W^2}{ M_Z^2\over (s-M_Z^2)^2}
%-------------------------
\\
&\times
\begin{cases}
%-------------------------
\left(1\pm \cos {\theta}\right)^2{g_e^-}^2+\left(1\mp \cos {\theta}\right)^2{g_e^+}^2,   \qquad\mbox{for $ \lambda=\pm1 $,}
%-------------------------
\\
%-------------------------
2\sin^2{\theta} \left({g_e^-}^2+{g_e^+}^2\right)\left(1+{\beta^2s\over 4M_Z^2}\right), \qquad\;\;\;\mbox{for $ \lambda=0 $,}
%-------------------------
\nn
\end{cases}
%-------------------------
\end{align}
%-------------------------
%-------------------------
with $\theta$ being the angle between ${\bf p}_Z$ and ${\bf k}_1$ in the CM frame,
$\beta={2|{\bf p_Z}|\over \sqrt{s}}$.
%By measuring the angular distribution of the $\mu^+\mu^-$ pair from $Z$ decay,
%it is possible to disentangle the different polarized cross sections experimentally.
Upon angular integration, the LO integrated cross section for polarized $Z$ reads:
%-------------------------
\bqa
%-------------------------
\label{Polarized:LO:X:section}
%-------------------------
&& \sigma_\lambda^{(0)}=\frac{\pi\alpha^2\beta\left({g_e^-}^2+{g_e^+}^2\right)}{6c_W^2 s_W^2}{M_Z^2\over (s-M_Z^2)^2}\left(1+\delta_{\lambda,0}{\beta^2 s \over 4M_Z^2}\right).
\nn\\
%-------------------------
\eqa
%-------------------------
%As can be readily seen from (\ref{LO:Differential:X:section}) and (\ref{Polarized:LO:X:section}).
The total unpolarized cross section $\sigma^{(0){\rm unpol}}=\sigma_L^{(0)}+\sigma_T^{(0)}\equiv
\sigma_0^{(0)}+2\sigma_{\pm 1}^{(0)}$.
In the high energy limit, the cross section for producing longitudinally-polarized $Z$ ($\propto 1/s$)
dominates the one associated with the transversely-polarized $Z$ ($\propto 1/s^2$),
%Nevertheless, at moderate energy, $\sqrt{s}= 250$ GeV,
%$\sigma_{L}^{(0)}$ only comprises of 42\% of the integrated unpolarized cross section.

\vspace{0.2 cm}
{\noindent \it \color{blue} The outline of calculation for radiative corrections.}
As far as the ${\cal O}(\alpha)+{\cal O}(\alpha\alpha_s)$ corrections are concerned, the higher-order diagrams can be grouped into
several distinct topologies as shown in Fig.~\ref{Fig:eeHZ:diagrams} and Fig.~\ref{Fig:SE:Vertex:diagrams}.

%First we comment on the Initial-State-Radiation (ISR) effect.

It is conventional to separate the ${\cal O}(\alpha)$ corrections into the electromagnetic and
weak corrections in a gauge-invariant manner.
%The electromagnetic corrections include the soft photon bremsstrahlung and the virtual photon exchange diagrams shown in
The NLO QED corrections as shown in Fig.~\ref{Fig:eeHZ:diagrams} are usually encoded in the so-called
Initial State Radiation (ISR) effect, which has been well-understood and implemented in Monte Carlo event generators.
A recent study using the package \textsf{WHIZARD}~\cite{Kilian:2007gr} reveals that,
including the ISR effect reduces the Born order $\sigma(HZ)$ at $\sqrt{s}=250$ GeV
by 10\%~\cite{Mo:2015mza}. A more careful analysis of the ISR effect for this process will be
presented elsewhere.

The ${\cal O}(\alpha)$ and ${\cal O}(\alpha\alpha_s)$  corrections to the amplitude can be decomposed as follows:
%-------------------------
\bqa
%-------------------------
&&\delta{\mathcal M}^{\sigma,\lambda} = \delta{\mathcal M}_{eeH}^{\sigma,\lambda}+
\delta{\mathcal M}_{\rm Box}^{\sigma,\lambda}+\delta{\mathcal M}_{eeZ}^{\sigma,\lambda}+\delta{\mathcal M}_{\rm S.E.}^{\sigma,\lambda}
%-------------------------
\nn\\
%-------------------------
&&+ \delta{\mathcal M}_{ZZH}^{\sigma,\lambda}
+ \delta{\mathcal M}_{\gamma ZH}^{\sigma,\lambda},
%-------------------------
\label{weak:corrections:assemble}
\eqa
%-------------------------
as can be recognized from Fig.~\ref{Fig:eeHZ:diagrams}.
The first two terms corresponding to the $eeH$ vertex corrections and box diagrams are UV-finite at ${\cal O}(\alpha)$.
%for the remaining diagrams we adopt the on-shell renormalization scheme~\cite{Sirlin:1980nh,Denner:1991kt} to deal with the %UV-divergences.

The amplitude arising from the $eeZ$ vertex corrections can be written as
$\delta {\mathcal{M}_{eeZ}^{\sigma,\lambda}}=\mathcal{M}_0^{\sigma,\lambda}\hat{\Gamma}^{\sigma}_{eeZ}$,
where the one-loop expression of the renormalized vertex form factor
$\hat{\Gamma}^{\sigma}_{eeZ}$ is given in \cite{Denner:1992bc}.
The amplitude also receives corrections from both $ZZ$ and mixed $\gamma Z$ self-energies:
%-------------------------
\bqa
%-------------------------
&& \delta\mathcal{M}_{\rm S.E.}^{\sigma,\lambda}=-\mathcal{M}_0^{\sigma,\lambda}
\left({\hat{\Sigma}_T^{ZZ}(s)\over s-M_Z^2}+{1\over g_e^{\sigma}}{\hat{\Sigma}_T^{\gamma Z}(s)\over s}\right),
%-------------------------
\label{corr:ampl:self:energy}
\eqa
%-------------------------
where $\hat{\Sigma}_T$ implies the renormalized transverse part of the gauge boson self-energy.

The amplitudes involving the $VZH(V=\gamma, Z)$ vertex corrections are
%-------------------------
\begin{subequations}\label{ZZH:gammaZH:corrections}
%-------------------------
\bqa
%-------------------------
%-------------------------
&&\delta\mathcal{M}_{ZZH}^{\sigma,\lambda}=\frac{e^2 g_e^{\sigma} M_Z}{s_W c_W}\bar{v}(k_1)
\gamma_{\mu}P_{\sigma}u(k_2){1\over s-M_Z^2}
\hat{\mathcal{T}}_{ZZH}^{\mu\nu}\varepsilon_{\lambda,\nu}^{*},
%-------------------------
\nn\\\\
%-------------------------
&& \delta\mathcal{M}_{\gamma Z H}^{\sigma,\lambda}=\frac{e^2 M_Z}{s_W c_W}\bar{v}(k_1)\gamma_{\mu} P_{\sigma} u(k_2){1\over s}\hat{\mathcal{T}}_{\gamma Z H}^{\mu\nu}\varepsilon_{\lambda,\nu}^{*}.
%-------------------------
\eqa
%-------------------------
\end{subequations}
%-------------------------
By Lorentz covariance, the vertex tensor $\hat{\mathcal{T}}_{VZH}^{\mu\nu}$ can be decomposed as
%-------------------------
\bqa
%-------------------------
&& \hat{\mathcal{T}}_{VZH}^{\mu\nu}=T_1 k^{\mu} k^{\nu}+T_2 p_Z^{\mu} p_Z^{\nu}+T_3
k^{\mu} p_Z^{\nu}+ T_4 p_Z^{\mu} k^{\nu}
%-------------------------
\nn\\
%-------------------------
&& +T_5 g^{\mu\nu}+T_6 \epsilon^{\mu\nu\rho\sigma} k_{\rho} p_{Z\sigma},
%-------------------------
\label{tensor:decomposition}
%-------------------------
\eqa
%-------------------------
where $k^{\mu}= p_{Z}^{\mu}+p_{H}^{\mu}$, and $T_i\;(i=1,\ldots,6)$
are Lorentz scalars solely depending on $s$, $M_H^2$ and $M_Z^2$.
%The hat in $\hat{\mathcal{T}}_{VZH}^{\mu\nu} $ means any possible UV divergence has been removed by adding the corresponding %counterterms.
%which can be find in \cite{Denner:1991kt}.
Among all form factors, only $T_5$ is subject to renormalization, and
the ${\cal O}(\alpha)$ counterterms for the $VZH$ ($V=Z,\gamma$) coupling
%-------------------------
%\begin{subequations}
%-------------------------
%\bqa
%-------------------------
%\delta^{\rm CT}_{ZZH}&=&\delta{Z_e}+{2s_W^2-c_W^2\over c_W^2}{\delta{s_W}\over s_W}+{1\over 2}{\delta{M_W^2}\over M_W^2}
%-------------------------
%\\
%-------------------------
%+{1\over
%2}\delta{Z_H}+\delta{Z_{ZZ}},
%-------------------------
%\\
%-------------------------
%\delta^{\rm CT}_{\gamma ZH} &=& \frac{1}{2}\delta{Z_{Z\gamma}}.
%-------------------------
%\eqa
%-------------------------
%\end{subequations}
%-------------------------
%The definitions of various renormalization constants above
can be found in \cite{Denner:1991kt}.
Beyond LO, the form factors $T_i(i=1,\cdots,5)$ do not vanish in general. Nevertheless,
due to Furry theorem, ${T}_6=0$ for both $ZZH$ and $\gamma ZH$ vertex corrections through ${\cal O}(\alpha\alpha_s)$.
Owing to the current conservation for massless electron, only $T_{4,5} $ contribute to the differential cross sections.

Some care should be exercised on the charge renormalization constant $Z_e$.
In the so-called $\alpha(0)$ scheme, where the $\alpha$ is assuming its Thomson-limit value,
$\delta Z_e$ can be expressed as
%-------------------------
$ \delta Z_e|_{\alpha(0)}= {1\over 2}{\Pi^{\gamma\gamma}(0)}-{s_W\over c_W}
{\Sigma^{\gamma Z}_T(0)\over M_Z^2}$,
%-------------------------
where ${\Pi(s)}\equiv {\Sigma_T^{\gamma\gamma}(s)\over s}$.
The first term in $\delta Z_e|_{\alpha(0)}$ is sensitive to the hadronic contribution,
thereby an intrinsic non-perturbative quantity. The hadronic contributions are often
absorbed into a non-perturbative parameter, $\Delta{\alpha_{\rm had}^{(5)}}(M_Z)$, which can be
extracted from the
measured $R$ values in low-energy $e^+e^-$ experiments~\cite{Olive:2016xmw}.
Equivalently, one can rewrite $\delta{Z_e}$ in $\alpha(0)$ scheme as
%-------------------------
\bqa
%-------------------------
\label{dZe:alpha(0):scheme}
%-------------------------
&& \delta{Z_e}\big|_{\alpha(0)}={1\over 2} \Delta{\alpha_{\rm had}^{(5)}}(M_Z)
+{1\over 2}{\rm Re}\, \Pi^{\gamma\gamma(5)}(M_Z^2)
%-------------------------
\nn\\
%-------------------------
&& +{1\over 2}\Pi^{\gamma\gamma}_{\rm rem}(0)-{s_W\over c_W}
{\Sigma^{\gamma Z}_T(0)\over M_Z^2},
%-------------------------
\eqa
%-------------------------
where $\Pi^{\gamma\gamma(5)}(M_Z^2)$ is the photon vacuum polarization from five massless quarks at momentum transfer
$M_Z^2$, and $\Pi^{\gamma\gamma}_{\rm rem}(0)$ represents the vacuum polarization from $W$ boson, charged leptons and top quark at zero momentum transfer.
Note these terms can be computed order by order in perturbation theory.
Throughout this work, we only retain the top quark mass and treat the
remaining five quarks massless (The effect of finite $m_b$ will be mentioned afterwards).

Two other popular parameterization schemes are the so-called $\alpha(M_Z)$ and $G_\mu$ schemes.
The corresponding charge renormalization constant can be converted from the $\alpha(0)$ scheme by
$\delta{Z_e}\big|_{\alpha(M_Z)}=\delta{Z_e}\big|_{\alpha(0)}- {1\over 2} \Delta\alpha(M_Z) $
and $ \delta Z_e|_{G_\mu}=\delta Z_e|_{\alpha(0)}-{1\over 2}\Delta r $ respectively,
%%-------------------------
%\begin{subequations}
%%-------------------------
%\bqa
%%-------------------------
%\label{dZe:alpha(MZ):scheme}
%%-------------------------
%\delta{Z_e}\big|_{\alpha(M_Z^2)}=&&\delta{Z_e}\big|_{\alpha(0)}- {1\over 2} \Delta\alpha(M_Z^2),
%%-------------------------
%\\
%%-------------------------
%\label{dZe:Gmu:scheme}
%%-------------------------
%\delta Z_e|_{G_\mu}=&&\delta Z_e|_{\alpha(0)}-{1\over 2}\Delta r,
%%-------------------------
%%-------------------------
%\eqa
%%-------------------------
%\end{subequations}
%%-------------------------
where $\Delta\alpha(M_Z)=\Pi^{\gamma\gamma}_{f\neq t}(0)-{\rm Re}\,\Pi^{\gamma\gamma}_{f\neq t}(M_Z^2)$,
and the expression for the oblique parameter $\Delta r $ can be found in
\cite{Denner:1991kt}.
The fine-structure constant can in turn be replaced with
%%-------------------------
\begin{subequations}
%%-------------------------
\bqa
%%-------------------------
\label{dZe:alpha(MZ):scheme}
%%-------------------------
&& \alpha\left(M_Z\right)=\frac{\alpha(0)}{1-\Delta \alpha\left(M_Z\right)},
%-------------------------
\\
%\text{and}
%\\
%-------------------------
&& \alpha_{G_\mu}=\frac{\sqrt{2}}{\pi}G_{\mu} M_W^2\left(1-\frac{M_W^2}{M_Z^2}\right)
%%-------------------------
\eqa
%%-------------------------
\end{subequations}
%%-------------------------
in the $\alpha(M_Z)$ and $G_\mu$ schemes, respectively. In contrast to the
$\alpha(0)$ scheme, these two schemes effectively resum some universal large (non-)logarithms
arising from the light fermions and top quark.

The ${\cal O}(\alpha\alpha_s^n)$ corrections to the differential cross section read
%-------------------------
\bqa
%-------------------------
&& {d \sigma_{\lambda}^{(\alpha\alpha_s^n)}\over d \cos\theta}=
{1\over4} {\beta\over 32\pi s} \sum_\sigma 2 {\rm Re}
\left[(\mathcal{M}_0^{\sigma,\lambda})^{*}\delta {\mathcal M}^{\sigma,\lambda}_{(\alpha\alpha_s^n)}\right],
%-------------------------
\label{corr:diff:cross:section}
\eqa
%-------------------------
where $n=0, 1$ represent the $ \mathcal{O}{(\alpha)} $  and $ \mathcal{O}{(\alpha\alpha_s)} $
corrections, respectively.

For the actual calculation, we work in Feynman gauge and adopt the dimensional regularization to regularize the UV divergences.
The Feynman diagrams and corresponding amplitudes are generated by \textsf{FeynArts}~\cite{Hahn:2000kx}.
The packages \textsf{FeynCalc/FormLink}~\cite{Mertig:1990an,Feng:2012tk} are employed to
carry out the trace over Dirac and color matrices, and the packages \textsf{Apart}~\cite{Feng:2012iq}
and \textsf{FIRE}~\cite{Smirnov:2014hma} are utilized
to perform partial fraction together with integration-by-parts (IBP) reduction.
We then combine \textsf{FIESTA}/\textsf{CubPack}~\cite{Smirnov:2013eza,CubPack} to perform
sector decomposition and subsequent numerical integrations for Master Integrals (MI)
with quadruple precision.

\vspace{0.2 cm}
{\noindent \it \color{blue} Next-to-leading order results.}
%First we revisit the NLO weak correction calculation for Higgs-strahlung process.
First we revisit the NLO weak corrections for the Higgsstrahlung process, in line with
(\ref{weak:corrections:assemble}) and (\ref{corr:diff:cross:section}).
We have worked out the bare NLO amplitude analytically and also employed \textsf{LoopTools}~\cite{Hahn:1998yk} for
an independent cross-check. After implementing various one-loop counterterms
analytically recorded in \cite{Denner:1991kt}, we have compared our UV-finite NLO predictions with numerous differential and integrated cross sections enumerated in \cite{Denner:1992bc}, and found gross agreement.
We have also compared our integrated NLO cross sections with those high-precision
predictions tabulated in \cite{Belanger:2003sd}, which utilized the automatic package
\textsf{GRACE-loop}. Reassuringly, for a variety of input values of $\sqrt{s}$ and $M_H$,
we always found better-than-per-mille agreement.

\vspace{0.2 cm}
{\noindent \it \color{blue} Mixed electroweak-QCD two-loop corrections.}\label{NNLO}
%-----------------------------
At ${\cal O}(\alpha\alpha_s)$, a simplifying pattern arises, {\it i.e.,}
the box diagrams and $eeH$ vertex are immune to gluonic dressing,
and only those two-loop diagrams of $s$-channel topology in Fig.~\ref{Fig:eeHZ:diagrams} survive.
Concretely, the mixed electroweak-QCD 2-loop corrections to the amplitude are expressed as the
last four terms in (\ref{weak:corrections:assemble}):
%-------------------------
%\bqa
%-------------------------
% \delta\mathcal{M}^{\sigma,\lambda(\alpha \alpha_s)}=\delta\mathcal{M}_{eeZ}^{\sigma,\lambda(\alpha \alpha_s)}+\delta\mathcal{M}_{\rm S.E.}^{\sigma,\lambda(\alpha \alpha_s)}
%-------------------------
%\nn\\
%-------------------------
%+\delta\mathcal{M}_{ZZH}^{\sigma,\lambda (\alpha \alpha_s)}+\delta\mathcal{M}_{\gamma ZH}^{\sigma,\lambda(\alpha \alpha_s)}.
%-------------------------
%\label{Mixed:EW-QCD:corr:to:ampl}
%-------------------------
%\eqa
%-------------------------
%Note there exists no factorizable ${\cal O}(\alpha)\otimes
%{\cal O}(\alpha_s)$ corrections that belong to two distinct parts.

As shown in Fig.~\ref{Fig:SE:Vertex:diagrams},  QCD renormalization is fulfilled by merely inserting the
top quark mass counterterm, $\delta m_t$, into the internal top quark propagator, as well as into the $Ht\bar{t}$ vertex.
We take $\delta m_t$ from~\cite{Djouadi:1993ss}:
%-------------------------
\beq
%-------------------------
\delta{m_t}=-m_t \Gamma(1+\epsilon)\left({4\pi \mu^2\over m_t^2}\right)^{\epsilon}{C_F\over 4}{\alpha_s\over \pi}{3-2\epsilon\over\epsilon(1-2\epsilon)},
%-------------------------
\eeq
%-------------------------
with the spacetime dimensions $d=4-2\epsilon$.

For the $\delta\mathcal{M}_{\rm S.E.}^{\sigma,\lambda(\alpha \alpha_s)}$ in
(\ref{corr:ampl:self:energy}), one can transplant the analytic ${\cal O}(\alpha\alpha_s)$ expressions
of the gauge boson/Higgs self-energies from \cite{Djouadi:1993ss,Djouadi:1994gf,Kniehl:1994ph},
and deduce the ${\cal O}(\alpha\alpha_s)$ corrections to
the renormalization constants $\delta Z_{e}$, $\delta Z_{\gamma Z}$, $\delta Z_{Z\gamma}$, $\delta Z_{ZZ}$, $\delta Z_H$, $\delta M_Z^2$, and $\delta M_W^2$.
Despite the absence of the ${\cal O}(\alpha\alpha_s)$ corrections to the bare $eeZ$ vertex,
one must incorporate the contribution to $\hat{\Gamma}^{\sigma(\alpha \alpha_s)}_{eeZ}$
that stems from the $\mathcal{O}(\alpha \alpha_s)$ counterterms,
$\delta^{{\rm CT}\pm (\alpha \alpha_s)}_{eeZ}$, which are UV-finite.
Their numerical values in the $\alpha(0)$ scheme are
%-------------------------
\begin{subequations}
%-------------------------
\label{counter:term:eeZ:alpha-alphas}
%-------------------------
\bqa
%-------------------------
&& \delta^{{\rm CT}+(\alpha \alpha_s)}_{eeZ}={\alpha \alpha_s \over \pi^2}\times (-27.33),
%-------------------------
\\
%-------------------------	
&&\delta^{{\rm CT}-(\alpha \alpha_s)}_{eeZ}={\alpha \alpha_s \over \pi^2}\times (+54.53).
%-------------------------
\eqa
%-------------------------
\end{subequations}
%-------------------------
The values enumerated in (\ref{counter:term:eeZ:alpha-alphas}) can be converted into the $G_\mu$-scheme
by subtracting $\tfrac{1}{2}
{\Delta r}^{(\alpha\alpha_s)}={\alpha \alpha_s \over \pi^2}\times (+22.49)$, which then agree with~\cite{Dittmaier:2014qza} when
adjusting the input parameters accordingly.

The real challenge is to compute the mixed electroweak-QCD corrections to $VZH$ vertex in
(\ref{ZZH:gammaZH:corrections}).
After IBP reduction, we end up with 47 MIs associated with
the bare two-loop diagrams, most of which involve four distinct scales.
Fortunately, at $\sqrt{s}\sim 250$ GeV, with the aid of \textsf{CubPack}~\cite{CubPack},
we can readily obtain very accurate results for all scalar form factors
$T_i\;(i=1,\cdots,5)$ in (\ref{tensor:decomposition}).

Ward identity for the $\gamma ZH$ vertex
%{\it i.e.},$k_{\mu}\hat{\mathcal{T}}_{\gamma ZH}^{\mu\nu}=0$,
demands $s {T}_{1}+k\cdot p_Z {T}_{4}+{T}_{5}=0$.
%-------------------------
We have numerically verified this relation to an extraordinary precision at ${\cal O}(\alpha\alpha_s)$.

Piecing together all the ${\cal O}(\alpha\alpha_s)$ ingredients,
we obtain the differential (un)polarized cross section following (\ref{corr:diff:cross:section}).
It is convenient to split the integrated (un)polarized cross sections into
%-------------------------
\beq
%-------------------------
\sigma_{\lambda}^{(\alpha \alpha_s)}=
\sigma_{\lambda,Z}^{(\alpha \alpha_s)}+\sigma_{\lambda,\gamma}^{(\alpha \alpha_s)}.
%-------------------------
\label{decompose:NNLO:pol:cross:section}
%-------------------------
\eeq
%-------------------------
For simplicity, we have combined the corrections originating from the $eeZ$ vertex, $ZZ$ self-energy and from the
$ZZH$ vertex together, dubbed $\sigma_{\lambda,Z}^{(\alpha \alpha_s)}$. Similarly,
$\sigma_{\lambda,\gamma}^{(\alpha \alpha_s)}$ is constructed
by merging the corrections from the $\gamma Z$ self-energy and from the
$\gamma ZH$ vertex.

\vspace{0.2 cm}
{\noindent \it \color{blue} Phenomenology.}
%-------------------------
We will take $\sqrt{s}=240,\:250$ GeV as two benchmark energy points at Higgs factory.
We adopt the following values for the input parameters~\cite{Olive:2016xmw}:
$M_H=125.09$ GeV, $M_Z=91.1876$ GeV, $M_W=80.385$ GeV, $m_t=174.2$ GeV,
$m_e=0.5109989$ MeV, $m_{\mu}=105.65837$ MeV, $m_{\tau}=1.77686$ GeV, $G_{\mu}=1.1663787\times 10^{-5}~{\rm GeV}^{-2}$, $\alpha(0)=1/137.035999$, $\Delta\alpha_{\rm had}^{(5)}(M_Z)=0.02764$ and $\alpha(M_Z)=1/128.943$ in the $\alpha(M_Z)$ scheme. We take $ \alpha_s(M_Z)=0.1185 $ as the initial value of the QCD running coupling and $\alpha_s(\mu)$ is evaluated with package \textsf{RunDec}~\cite{Chetyrkin:2000yt}.

%-------------------------
\begin{table*}[htbp]
	\begin{tabular}{|c|c|c|c|c|c|c|c|c|}
		\hline
		\multirow{2}{*}{$ \sqrt{s} $ (GeV)}& &LO (fb) & \multicolumn{2}{c|}{NLO Weak (fb)} & \multicolumn{4}{c|}{NNLO mixed electroweak-QCD (fb)}
		\\
		\cline{3-9}
		\multicolumn{1}{|c|}{}&& $\sigma^{(0)}$ & $\sigma^{(\alpha)}$ & $\sigma^{(0)} +\sigma^{(\alpha)}$ &  $\sigma_{Z}^{(\alpha\alpha_s)}$ & $\sigma_{\gamma}^{(\alpha\alpha_s)}$ & $\sigma^{(\alpha\alpha_s)}$ &  $ \sigma^{(0)}+\sigma^{(\alpha)}+\sigma^{(\alpha \alpha_s)} $ \\
		\hline
		\multicolumn{1}{|c|}{}&{Total}& {\bf 223.14} & $6.64$&  {\color{blue}$229.78$}  & $2.42$ & $0.008$ & $2.43$ &  {\color{red} $232.21$}
		\\
		\cline{2-9}
		\multicolumn{1}{|c|}{$240$}&{L}&88.67&$3.18$& $91.86$   & $0.96$ & $0.003$ & $0.97$ & $92.82$
		\\
		\cline{2-9}
		\multicolumn{1}{|c|}{}&{T}&134.46&$3.46$&   $137.92$   & $1.46$ & $0.005$ & $1.46$ & $139.39$
		\\
		\hline
		\multicolumn{1}{|c|}{}&{Total}& {\bf 223.12} & $6.08$&  {\color{blue} $229.20$}  & $2.42$ & $0.009$ & $2.42$ & {\color{red}$231.63$}
		\\
		\cline{2-9}
		\multicolumn{1}{|c|}{$250$}&{L}&94.30&$3.31$&   $97.61$  & $1.02$ & $0.004$ & $1.02$ & $98.64$
		\\
		\cline{2-9}
		\multicolumn{1}{|c|}{}&{T}&128.82&$2.77$&  $131.59$  & $1.40$ & $0.005$ & $1.40$ & $132.99$
		\\
		\hline
	\end{tabular}
	\caption{The (un)polarized Higgsstrahlung cross sections at $\sqrt{s}=240$ GeV and $250$ GeV in the $\alpha(0)$ scheme.
Provided are the LO, NLO weak and NNLO $\mathcal{O}(\alpha \alpha_s)$ predictions as well as
individual contributions for the $\mathcal{O}(\alpha)$ corrections $\sigma^{(\alpha)}$,
and for the $\mathcal{O}(\alpha \alpha_s)$ corrections in (\ref{decompose:NNLO:pol:cross:section}).
\label{Table:1}
}
\end{table*}
%-------------------------

Table~\ref{Table:1} lists our LO, NLO, and NNLO predictions to the integrated (un)polarized Higgsstrahlung cross sections
in the $\alpha(0)$ scheme.
While the unpolarized cross sections at $\sqrt{s}=240,\,250$ GeV are quite close in magnitude,
$\sigma_L$($\sigma_T$) are slightly bigger(smaller) in the case of the higher energy.
The NLO weak corrections increase the $\sigma^{(0)}$ by 3.0\%(2.7\%) at $\sqrt{s}=240(250)$ GeV.
The NNLO electroweak-QCD corrections turn out to be sizable, about 1.1\% of the LO cross section for both CM energies.

One interesting feature can be recognized from Table~\ref{Table:1}, the
$\sigma_{\lambda,\gamma}^{(\alpha \alpha_s)}$
in (\ref{decompose:NNLO:pol:cross:section}) turns out to be much suppressed.
This is compatible with the tiny ${\cal O}(\alpha\alpha_s)$ corrections to $H\to Z\gamma$ found in \cite{Spira:1991tj,Gehrmann:2015dua,Bonciani:2015eua}.

%-------------------------
\begin{table}[htbp]
	\renewcommand\arraystretch{1.2}
	\begin{tabular}{|c|c|c|c|c|}
		%\hline
		%\multirow{2}{*}{$ \sqrt{s} $ }  & \multicolumn{3}{c|}{$\alpha(0)$ {\rm scheme}} & \multicolumn{3}{c|}{$\alpha(M_z)$ {\rm scheme}} & \multicolumn{3}{c|}{$G_\mu$ {\rm scheme}}\\
		%\cline{2-10}
		\hline
		$ \sqrt{s} $ & schemes &  $\sigma_{\rm LO}$ (fb) & $\sigma_{\rm NLO}$ (fb) & $\sigma_{\rm NNLO}$ (fb)
		\\
		\hline
		&$ \alpha(0) $& $223.14 \pm 0.47$ & $229.78 \pm 0.77$ & $232.21_{-0.75-0.21}^{+0.75+0.10}$
		\\
		\cline{2-5}
		240&$ \alpha(M_Z) $&$252.03 \pm 0.60$  & $228.36_{-0.81}^{+0.82}$ & $231.28_{-0.79-0.25}^{+0.80+0.12}$
		\\
		\cline{2-5}
		&$ G_{\mu} $&$239.64 \pm 0.06$ & $232.46_{-0.07}^{+0.07}$ & $233.29_{-0.06-0.07}^{+0.07+0.03}$
		\\
		\hline
		\hline
		&$ \alpha(0) $& $223.12 \pm 0.47$ & $229.20 \pm 0.77$ & $231.63_{-0.75-0.21}^{+0.75+0.12}$
		\\
		\cline{2-5}
		250&$ \alpha(M_Z) $& $252.01 \pm 0.60$ & $227.67_{-0.81}^{+0.82}$ & $230.58_{-0.79-0.25}^{+0.80+0.14}$
		\\
		\cline{2-5}
		&$ G_{\mu} $& $239.62 \pm 0.06$ & $231.82{\pm0.07}$ & $232.65_{-0.07-0.07}^{+0.07+0.04}$
		\\
		\hline
	\end{tabular}
	\caption{The unpolarized Higgsstrahlung cross sections at $\sqrt{s}=240 (250)$ GeV in
three different input schemes. To estimate the uncertainties caused by the input parameters (first entry),
		we take $M_W=80.385 \pm 0.015\,{\rm GeV}$, $m_t=174.2 \pm 1.4\,{\rm GeV}$ and $\Delta\alpha_{\rm had}^{(5)}(M_Z)=0.02764\pm 0.00013$.
We also change the strong coupling constant from $\alpha_s(M_Z)$ to $\alpha_s(\sqrt{s})$ (second entry) with
its central value taken as $\alpha_s=\alpha_s(\sqrt{s}/2)$.
For the conversion from the $\alpha(0)$ scheme to the $\alpha(M_Z)$ and $G_{\mu}$
schemes, we use $\Delta \alpha(M_Z)|_{\text{NLO}}=\Delta \alpha(M_Z)|_{\text{NNLO}}=0.059$
and $\Delta r|_{\text{NLO}}=0.0293, \Delta r|_{\text{NNLO}}=0.0331$, respectively.}
	\label{Table:2}
\end{table}

In Table II we provide our LO, NLO, NNLO predictions for the
unpolarized Higgsstrahlung cross sections in the three input schemes
together with the parametric uncertainty (first entry) and the QCD
renormalization scale uncertainty (second entry).
To assess the parametric uncertainty, we vary the values of $M_W$ and $m_t$, $\Delta\alpha_{\rm had}^{(5)}(M_Z)$
within the PDG-quoted $1-\sigma$ error bands. For the QCD scale uncertainty,
we vary the the renormalization scale $\mu$ in $\alpha_s$
from $M_Z$ to $\sqrt{s}$.

While the parametric and scale uncertainties of the NNLO predictions
in the $\alpha(0)$ and $\alpha(M_Z)$ schemes are at the level of 0.3\% and 0.4\%
of the NNLO result, respectively, they
are considerably reduced in the $G_\mu$ scheme ($\approx$ 0.04\%). We also
find that in the $G_\mu$ scheme the NNLO electroweak-QCD corrections only
amount to 0.3\% of $\sigma^{(0)}$, which is due to the fact in addition to
the running of $\alpha$, universal corrections to the $\rho$ parameter are
also absorbed into the LO cross section. As can also be seen in Table~\ref{Table:2},
the sensitivity to the choice of input scheme is reduced at NNLO compared to NLO. To further reduce the input scheme dependence, one may have to include
the two-loop electroweak corrections as well.

%-------------------------
\begin{figure}[htbp]
	\centering
	\includegraphics[width=0.4\textwidth]{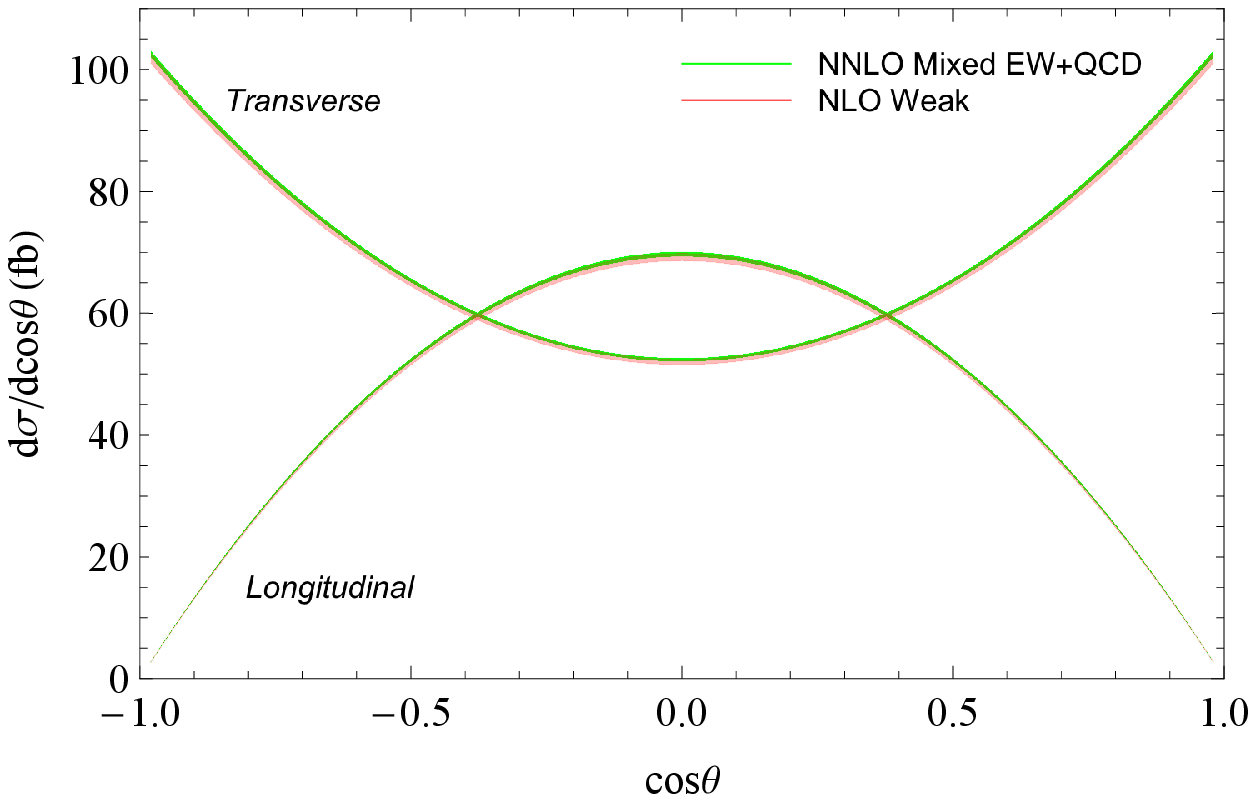}
	\includegraphics[width=0.4\textwidth]{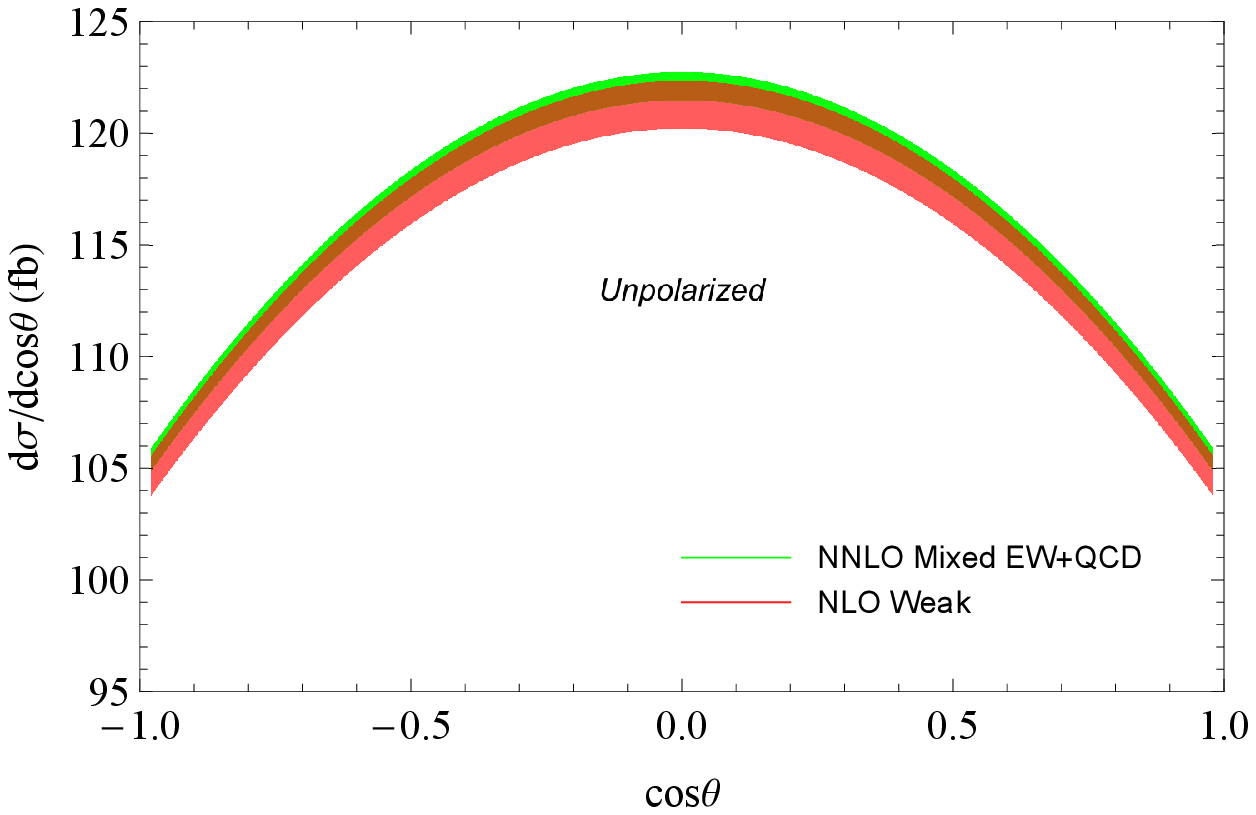}
	\caption{Differential unpolarized/polarized cross sections for Higgsstrahlung at $\sqrt{s}=240$ GeV
at NLO $ \mathcal{O}(\alpha) $ and NNLO $ \mathcal{O}(\alpha\alpha_s)$. The green band indicates the uncertainties from
the input parameters as adopted in Table~\ref{Table:2} and the three different input schemes.
		\label{Figures:diff:distribution:eeZH}}
\end{figure}
%-------------------------

In Fig.~\ref{Figures:diff:distribution:eeZH} we show the angular distribution of (un-)polarized $Z$
boson in $HZ$ production at a Higgs factory CM energy of $240$ GeV
at various levels of accuracy.

In our calculation, we neglected all quark masses except the top quark mass,
and thus the $b$ quark does not contribute to the $VHZ$ vertex diagram.
To access the validity of this approximation, we re-did our NLO and NNLO calculations by retaining $m_b=4.66$ GeV.
Due to the occurrence of the hierarchy $m_b\ll \sqrt{s}\sim M_H\sim M_Z$, this turns out to be a rather
challenging calculation. We find that, keeping finite $m_b$ reduces the NLO cross section at $\sqrt{s}=250$ GeV by 0.05 fb,
and reduces the final NNLO prediction by roughly 0.01 fb in the $\alpha(0)$ scheme.
This small impact of a finite bottom quark mass is completely overwhelmed by the uncertainties
listed in Table~\ref{Table:2}.

%The differential cross sections for (un)polarized $Z$ at various level of perturbative accuracy are
%shown in Fig.~\ref{Figures:diff:distribution:eeZH}.

\vspace{0.2 cm}
{\noindent \it \color{blue}Summary and Outlook.}
Stimulated by the anticipated exquisite accuracy of the $\sigma(HZ)$ measurements in the next-generation $e^+e^-$ Higgs factory,
for the first time we calculated the mixed electroweak-QCD ${\cal O}(\alpha\alpha_s)$ corrections for the Higgsstrahlung
process. It is found that this mixed electroweak-QCD corrections are quite sizable, about 1.1\% of the LO result in $\alpha(0)$ and $\alpha(M_Z)$ schemes, well above the projected experimental (sub-)percent accuracy for the $\sigma(ZH)$ measurement.
In the $G_\mu$ scheme, we find that the NNLO electroweak-QCD corrections
amount to 0.3\% of the LO result.
A comprehensive study of parametric and QCD scale uncertainties
exhibits large uncertainties in the NNLO electroweak-QCD predictions
in the $\alpha(0)$ and $\alpha(M_Z)$ schemes, which however are considerably
reduced in the $G_\mu$ scheme.
It is important to note that to make
closer contact with the actual experimental measurement, it is also useful to conduct a careful analysis on the
ISR effects, as well as to study the process $e^+e^-\to \mu^+\mu^-+H$ by including the effect of finite $Z$ width.

\vspace{0.2 cm}
{\noindent \it \color{blue} Note added.}
%-------------------------
After this work was submitted, there also appeared an independent computation on mixed electroweak-QCD corrections to
Higgsstrahlung process~\cite{Gong:2016jys}.
%Their results are compatible with ours in the $\alpha(0)$ and $\alpha(M_Z^2)$
%schemes if identical input parameters are chosen. Moreover, we have also included the predictions from the $G_\mu$ scheme, %investigated the nonzero $m_b$ effect, and performed a comprehensive uncertainty analysis by combining three
%parameterization schemes.

\vspace{0.2 cm}
\begin{acknowledgments}
{\noindent\it \color{blue} Acknowledgment.}
%-----------------------
We are grateful to Gang Li, Xiaohui Liu and Jian-Hui Zhang for useful discussions.
%-----------------------
Q.-F.~S. wishes to thank Theory Division of IHEP for warm hospitality,
where this work was being finalized.
%-----------------------
Q.-F.~S. is supported by the National Natural Science Foundation of China under Grant No. 11375168 and No.~11475188.
%-----------------------
The work of F.~F. is supported by the National Natural Science Foundation of China under Grant No.~11505285,
and by the Fundamental Research Funds for the Central Universities.
%-----------------------
The work of Y.~J. is supported in part by the National Natural Science Foundation of China under Grants
No.~11475188, No.~11261130311, No.~11621131001 (CRC110 by DGF and NSFC), by the IHEP Innovation Grant under contract number Y4545170Y2,
and by the State Key Lab for Electronics and Particle Detectors.
%-----------------------
W.-L.~S. is supported by the National Natural Science Foundation of China under Grant No.~11447031 and No. 11605144, by the Natural
Science Foundation of ChongQing under Grant No. cstc2014jcyjA00029,
and also by the Fundamental Research Funds for the Central Universities under Grant No. XDJK2016C067.
The Feynman diagrams in this paper were prepared using JaxoDraw~\cite{Binosi:2003yf,Binosi:2008ig}.
%-----------------------
\end{acknowledgments}

\end{document}